\documentclass[prl,twocolumn,amsmath,amssymb,superscriptaddress]{revtex4}
\usepackage{graphicx}
\usepackage{dcolumn}
\usepackage{bm}

\begin{document}

\title{Phonon switching and combined Fano-Rice effect
in optical spectra of bilayer graphene}

\author{E. Cappelluti}
\affiliation{Istituto dei Sistemi Complessi, sezione Sapienza, CNR,
via dei Taurini 19, 00185 Roma, Italy}
\affiliation{Dipartimento di Fisica, Universit\`a ``La Sapienza'',
P.le A. Moro 2, 00185 Rome, Italy}

\author{L. Benfatto}
\affiliation{Istituto dei Sistemi Complessi, sezione Sapienza, CNR,
via dei Taurini 19, 00185 Roma, Italy}
\affiliation{Dipartimento di Fisica, Universit\`a ``La Sapienza'',
P.le A. Moro 2, 00185 Rome, Italy}

\author{A.B. Kuzmenko}
\affiliation{DPMC,
Universit\'{e} de Gen\`{e}ve, 1211 Gen\`{e}ve,
Switzerland}

\begin{abstract}
Recent infrared measurements of phonon peaks in gated bilayer graphene
reveal two striking signatures of electron-phonon interaction: an
asymmetric Fano lineshape and a giant variation of the peak intensity as a
function of the applied gate voltage.  In this Letter we provide a unified
theoretical framework which accounts for both these effects and unveils
the occurrence of a switching mechanism between the
symmetric ($E_g$) and anti-symmetric ($E_u$) phonon mode
as dominant channel in the optical response.
A complete phase diagram of the optical phonon response is also
presented, as a function of both the charge density and the bandgap.
\end{abstract}

\date{\today}

\maketitle

Single  and multi-layer graphenes are among the most promising
systems for the development of carbon-based devices in electronics.
Bilayer graphene is of particular interest
because a controlled tunable gap in the electronic spectrum
can be there induced by applying one (or more) external gate
voltages \cite{McCann2,Castro,Oostinga},
as observed by transport \cite{Oostinga} and optical
measurements \cite{Zhang,Mak,Kuzmenko1,Kuzmenko2}.
The potential interest in application has triggered also an intense research
on the vibrational properties, that can be used 
for instance for a careful characterization of the number of layers,
the charge doping and the amount of disorder \cite{Stampfer,Ferrari,Yan07}.
Large part of the work in this context
has focused on the properties of the in-plane $E_g$ mode
at $\omega \approx 0.2$ eV
which is present in single as well as in bilayer
systems, and which can be probed by Raman
spectroscopy \cite{Ferrari,Pisana,Yan07,Yan_PRL,Malard,Kim}.

Recently, phonon peaks in the energy range $\omega \approx 0.2$ eV have
been reported also in infrared (IR) optical measurements of bilayer
graphene \cite{noi,Tang}, showing a rather different phenomenology with
respect to Raman spectroscopy.  In particular, the observed IR phonon peak
presents a strong dependence of the intensity and of the Fano-like
asymmetry as a function of the applied gate voltage \cite{noi,Tang}.  These
features have been attributed respectively to a charged-phonon effect for
the $E_u$ antisymmetric (A) mode \cite{noi}, or, at $n=0$, to the emergence
of a Fano profile for the $E_g$ symmetric (S) mode, whose optical activity
can be triggered by the electrostatic potential difference $\Delta$
between the two carbon planes \cite{Tang}.  However, the possible
connection between these two alternative views is still unclear.

In this Letter we provide a unified microscopic framework that allows us to
elucidate the relative role of the $E_u$ and $E_g$ phonon modes in bilayer
graphene, with regards to the infrared activity and the Fano asymmetry of the
observed phonon peaks.  We present a complete phase diagram for the
strength of the phonon modes and their Fano properties as functions of the
chemical potential and $\Delta$, showing that a switching mechanism between
the dominance of the $E_u$ or $E_g$ mode can be controlled by the external
gate voltage. Our work permits thus reconciling within a unique approach
the phonon-peak features observed by different experimental groups
\cite{noi,Tang}.

\begin{figure}[b]
\includegraphics[scale=0.37,clip=]{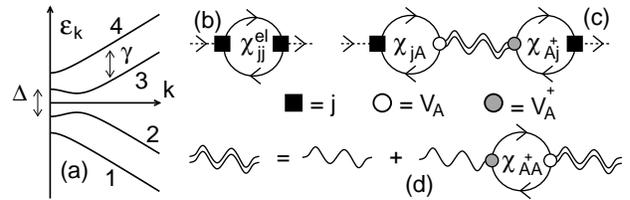}
\caption{(a) Scheme of the band structure.
(b)-(d) Relevant diagrams entering the optical conductivity for $\Delta=0$:
Dashed, solid and wavy
lines represent the photon, the electron and phonon Green's functions,
respectively.
Squares and circles
are the current and the electron-phonon scattering matrices 
$\hat{j}$, $\hat{V}_{\rm A}$, $\hat{V}_{\rm A}^\dagger$, respectively.}
\label{f-diagrams}
\end{figure}

To compute the conductivity of bilayer graphene we 
work in the $4 \times 4$ basis of the atomic orbitals
$\Psi^\dagger_{\bf k}= (a_{1{\bf k}}^\dagger,b_{1{\bf k}}^\dagger,a_{2{\bf
k}}^\dagger, b_{2{\bf k}}^\dagger)$, where $a_{i \bf k}^\dagger$ and $b_{i
\bf k}^\dagger$ operators create an electron in the layer $i$ and on the
sublattice A or B, respectively.
In this basis, the Hamiltonian for bilayer graphene
near the K point reads
\cite{Ando_07}: $\hat{H}_{\bf k}= \left\{ \hbar v {\bf
k}\cdot\hat{I}(\hat{\mbox{\boldmath$\sigma$}})
+(\Delta/2)\hat{\sigma}_z(\hat{I})+ (\gamma/2) \left[
\hat{\sigma}_x(\hat{\sigma}_x) +\hat{\sigma}_y(\hat{\sigma}_y)
\right]\right\} $, where $\hat{\sigma}_i$ and $\hat{I}$ are $2\times 2$
Pauli matrices and the unit matrix respectively, and
$\hat{A}(\hat{B})\equiv \hat{A}\otimes \hat{B}$.  Here $v$ is the Fermi
velocity for single-layer graphene and $\gamma$ is the interlayer hopping.
The electrostatic potential difference $\Delta$ induces a gap in the
diagonalized bands $\epsilon_{{\bf k},n}$ \cite{McCann2}, labeled according
to Fig. \ref{f-diagrams}a.
%
%
%
%
We set the electric field of the infrared radiation along the $x$-axis, so
that the electric current reads $j=\sum_{\bf k} \Psi^\dagger_{\bf
k}\hat{j}_{\bf k}\Psi_{\bf k}$ where $\hat{j}_{\bf k}=-e v
\hat{I}(\hat{\sigma}_x)$.  The two $E_u$ and $E_g$ in-plane optical phonons
have degenerate longitudinal and transverse polarization at zero momentum
${\bf q}=0$ \cite{Ando_07}. We can write the electron-phonon interaction
for these modes as: ${H}_{\rm ep}= \sum_\nu V_\nu \phi_\nu$, where
$\phi_\nu$ is the dimensionless lattice displacement for the $\nu=$A,S
branch at ${\bf q}=0$, $V_{\nu}= \sum_{\bf k}, \Psi^\dagger_{\bf k}
\hat{V}_\nu \Psi_{\bf k}$ is the corresponding electron-phonon scattering
operator.  For a choice of the longitudinal polarization along $x$, one
finds $\hat{V}_{\rm A}=ig\hat{\sigma}_z(\hat{\sigma}_y)$ and $\hat{V}_{\rm
S}=ig\hat{I}(\hat{\sigma}_y)$ \cite{Ando_07}, where $g$ is the
electron-phonon coupling.

Let us consider first the case of ungapped bilayer graphene ($\Delta=0$),
where only the $A$ mode is optically active.  Since graphene is a non-polar
system, the bare dipole induced by the rigid shift of the valence charges
upon the $A$ lattice distortion is extremely small \cite{noi}.
Nevertheless, an optical phonon response can be still mediated by the
conduction charges in the presence of the electron-phonon interaction, as
suggested by M.J. Rice in the context of organic and fullerene compounds
\cite{Rice}.  The total optical conductivity of the system is computed as
$\sigma(\omega)=-\chi_{jj}(\omega)/i\omega$, where
$\chi_{jj}(\omega)=-\langle {j}{j}\rangle_\omega$ is the current-current
response function. In the presence of electron-phonon interaction, one can
identify two classes of contributions in $\chi_{jj}(\omega)$, namely
$\chi_{jj}(\omega)= \chi_{jj}^{\rm el}(\omega)+\chi_{jj}^{\rm ep}(\omega)$.
The former, depicted in Fig. \ref{f-diagrams}b, describes electronic
excitations (see, for example, Ref. \cite{Nicol}), while the latter
contains all the diagrams which can be split in two by cutting one phonon
propagator (Fig. \ref{f-diagrams}c) \cite{Rice}.  We can write thus
\begin{eqnarray}
\chi_{jj}^{\rm ep}(\omega)
&=&
\chi_{j{\rm A}}(\omega)D_{\rm AA}(\omega)\chi_{{\rm A}^\dagger j}(\omega),
\label{chirice}
\end{eqnarray}
where $\chi_{j{\rm A}}(\omega)=-\langle {j} V_{\rm A}\rangle_\omega$,
$\chi_{{\rm A}^\dagger j}(\omega)=-\langle V^\dagger_{\rm A}
j\rangle_\omega$ are the mixed current-phonon response functions, and
$D_{\rm AA}(\omega)= -\langle \phi_{\rm A} \phi_{\rm A}\rangle_\omega
\approx [\omega-\omega_{\rm A}+i\Gamma_{\rm A}/2]^{-1}$ is the phonon
propagator with frequency $\omega_{\rm A}$ and linewidth $\Gamma_{\rm A}$
renormalized by the phonon self-energy $\chi_{\rm A^\dagger A}$
(Fig. \ref{f-diagrams}d).  It should be emphasized that two {\em different}
response functions, $\chi^{\rm el}_{jj}$ and $\chi_{j{\rm A}}$, enter in
the above decomposition of $\sigma(\omega)$.  This distinction, that has
been neglected in the original formulation \cite{Rice}, is however crucial,
because it implies that the allowed particle-hole excitations of the system
will contribute in a different way to the electronic optical background,
related to $\chi^{\rm el}_{jj}$, to the phonon renormalization, controlled
by $\chi_{\rm A^\dagger A}$, or to the electron-phonon optical response,
controlled by $\chi_{j{\rm A}}$.

Eq. (\ref{chirice}) leads to the onset, in the real part of the optical
conductivity $\sigma'_{\rm ep}(\omega)=-{\rm Im}\chi_{jj}^{\rm
ep}(\omega)/\omega$, of a phonon peak at $\omega_{\rm A}$.  Indeed, using
the relation $\chi_{{\rm A}^\dagger j}(\omega)=\chi_{j{\rm A}}(\omega)$,
and taking the real and imaginary part of each element in Eq.\
(\ref{chirice}), we get
\begin{eqnarray}
\sigma'_{\rm ep}(\omega)\Big|_{\omega\approx \omega_{\rm A}}
&\approx&
\frac{2 \left[\chi_{j{\rm A}}'(\omega_{\rm A})\right]^2}
{\omega_{\rm A}\Gamma_{\rm A}}
\left[
\frac{q_{\rm A}^2-1+2q_{\rm A} z}{q_{\rm A}^2(1+z^2)}
\right],
\label{fanorice1}
\end{eqnarray}
where $z=2[\omega-\omega_{\rm A}]/\Gamma_{\rm A}$ and where
\begin{eqnarray}
q_{\rm A}
&=&
-\frac{\chi_{j{\rm A}}'(\omega_{\rm A})}{\chi_{j{\rm A}}''(\omega_{\rm A})}.
\label{fanorice2}
\end{eqnarray}
Eq. (\ref{fanorice1}) has exactly the same structure as the Fano formula
\cite{Fano}.  The derivation of Eqs. (\ref{fanorice1})-(\ref{fanorice2})
shows not only that in bilayer graphene the Fano effect originates from a
correct implementation of the charged-phonon Rice theory, but it
provides a compelling procedure to evaluate on microscopic grounds 
the shape and the intensity of the phonon peak. For instance from
Eqs. (\ref{fanorice1})-(\ref{fanorice2}) we can identify the
$\omega$-integrated peak area as $W'_{\rm A} = \pi[\chi_{j{\rm
A}}'^2(\omega_{\rm A})-\chi_{j{\rm A}}''^2(\omega_{\rm A})] /\omega_{\rm
A}=(1-1/q_{\rm A}^2)[\pi\chi_{j{\rm A}}'^2(\omega_{\rm A})/\omega_{\rm
A}]$. Since the sign of $W'_{\rm A}$ depends on $q_{\rm A}$ it can be
convenient, as done in Ref.\ \cite{noi}, to define a ``bare'' intensity as
$W_{\rm A} = \pi\chi_{j{\rm A}}'^2(\omega_{\rm A})/\omega_{\rm A}$, which
coincides with $W_{\rm A}'$ in the limit $|q_{\rm A}|\rightarrow \infty$,
when Eq.\ (\ref{fanorice1}) reduces to a conventional Lorentzian peak with
weight $W_{\rm A}$.  Note however that in the opposite case $|q_{\rm
A}|\approx 0$ one recovers from Eqs. (\ref{fanorice1})-(\ref{fanorice2}) a
completely {\em negative} Lorentzian peak, of intensity $-\pi
|\chi''_{j{\rm A}}(\omega_{\rm A})|^2/\omega_{\rm A}$, so that the
definition of `intensity' of the phonon peak can be ambiguous in the
presence of the Fano effect.  A more convenient quantity to parameterize
the strength of the phonon peak is $p_{\rm A}= \pi|\chi_{j{\rm
A}}(\omega_{\rm A})|^2/\omega_{\rm A} =W_{\rm A}(1+1/q_{\rm A}^2)$, which
is always positive and vanishes when the phonon peak is completely absent.

\begin{figure}[t]
\includegraphics[scale=0.4,clip=]{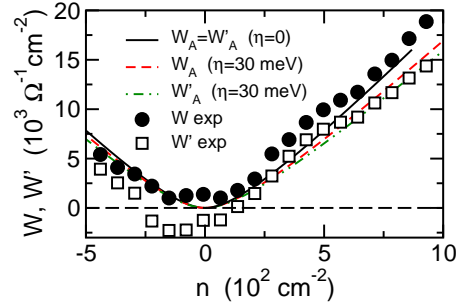}
\caption{(Color online) $W_{\rm A}$, $W_{\rm A}'$ intensities of the $E_u$
mode as a function of the charge concentration $n$ for $\Delta=0$
and $\eta=0,30$ meV.  Also shown are experimental data from Ref.\
\cite{noi} for the bare intensity $W_{\rm exp}$ and the $\omega$-integrated
one $W'_{\rm exp}$.}
\label{f-x}
\end{figure}

Eqs.\ (\ref{fanorice1})-(\ref{fanorice2}) can be computed analytically by
using the non-interacting electron Green's functions.  The quantity
$\chi_{jA}$ has the typical structure of a particle-hole Lindhard response
function, with proper coherence factors $C^{nm}_{jA}$ weighting the
contributions of the various excitations between the $n$ and
$m$ bands. In particular, using the explicit
matrix expressions of the $\hat j$ and $\hat V_A$ operators, one gets 
\begin{eqnarray}
\chi_{j{\rm A}}(\omega)
&=&
\chi_{j{\rm A}}^{12}(\omega)+\chi_{j{\rm A}}^{13}(\omega)
-\chi_{j{\rm A}}^{24}(\omega)-\chi_{j{\rm A}}^{34}(\omega),
\label{chimix}
\end{eqnarray}
where $\chi^{nm}_{jA}(\omega)=\pi^{nm}_{jA}(\omega)-
\pi^{mn}_{jA}(\omega)$, and 
\begin{eqnarray}
\pi_{j{\rm A}}^{nm}(\omega)
&=&
\sum_{\bf k}
 C_{j{\rm A},{\bf k}}^{nm}
\frac{f(\epsilon_{{\bf k},n}-\mu)-f(\epsilon_{{\bf k},m}-\mu)}
{\epsilon_{{\bf k},n}-\epsilon_{{\bf k},m}+\hbar\omega+i\eta}.
\label{elements}
\end{eqnarray}
Here $f(x)=1/[\exp(x/T)+1]$ is the Fermi function,
$C_{j{\rm A},{\bf k}}^{nm}=(gevN_s N_v)
\gamma/4\sqrt{(\hbar v k)^2+\gamma^2}$ for $n,m$ as in Eq.\ (\ref{chimix})
and zero otherwise, $\mu$ is the chemical potential,
$N_s=N_v=2$ are the spin and valley degeneracies, and
the $\eta$ factor takes into account broadening effects due to impurities
and inhomogeneities. In the clean limit, $\eta=0$,
we obtain the analytical expressions valid for $T=0$ and $|\mu| < \gamma$:
\begin{eqnarray}
\chi_{j{\rm A}}'(\omega)
&=&
A
\left\{
\ln\left[\frac{(1+u)(1-u+2w)}{(1-u)(1+u+2w)}\right]
+\frac{4uw}{(1-u^2)}
\right\}.
\label{rechi}
\\
\chi_{j{\rm A}}''(\omega)
&=&
\pi A
\left[
\theta(|u|-1)\theta(1-|u|+2w)+2w\delta(|u|-1)
\right],
\label{imchi}
\end{eqnarray}
where $A=ge\gamma/4\pi\hbar v$,
$u=\hbar\omega/\gamma$ and $w=|\mu|/\gamma$.  Similar analytical
expressions can be obtained for $\mu > \gamma$. 

To compute the optical conductivity (\ref{fanorice1}) we evaluate the
complex function $\chi_{j{\rm A}}(\omega)$ at the phonon resonance
$\omega_{\rm A}\approx 0.2$ eV, using standard values for $\hbar v=6.74$
eV\AA \, and $\gamma=0.39$ eV.  From the value $\alpha=6.4$ eV \AA$^{-1}$
\cite{CNG} of the deformation potential we get $g=0.27$ eV, corresponding
to the dimensionless parameter $\lambda=(\sqrt{3}/\pi)g^2/(\hbar v/a)^2=
6\times10^{-3}$ \cite{Ando_07}, in agreement with the experimental
estimates given in Refs. \cite{Malard,noi}.  The resulting spectral weight
$W_{\rm A}$ as evaluated from Eq.\ (\ref{fanorice1}) and
Eqs. (\ref{chimix})-(\ref{elements}) is shown in Fig. \ref{f-x} for
$\eta=0$ and $\eta=30$ meV, along with the experimental data $W_{\rm exp}$,
$W'_{\rm exp}$ taken from Ref. \cite{noi}.  Note that the only possible
electron-hole excitations at the phonon energy $\omega_{\rm A}$ ($<
\gamma$), namely the 2-3 multiband transitions, are not allowed in
Eq. (\ref{chimix}).  Thus in the clean limit $\chi_{j{\rm A}}''(\omega_{\rm
A})\approx 0$, so that no Fano effect ($q_{\rm A}=-\infty$) is found and
$W_{\rm A}'=W_{\rm A}$, while for $\eta\neq 0$ a small contribution to
$\chi_{j{\rm A}}''(\omega_{\rm A})$ from the other interband transitions
gives rise to a weak Fano asymmetry, and hence $W_{\rm A}'\lesssim W_{\rm
A}$, as shown in Fig.\ \ref{f-x}.  Both the magnitude of $W_A$, which
is proportional to the dimensionless coupling $\lambda$, 
and its doping dependence are in excellent agreement with $W_{\rm
exp}$, pointing out that the $E_u$ mode is the main responsible for the
phonon infrared intensity reported in Ref.\ \cite{noi} at large $n$.  On
the other hand, even in the presence of a large $\eta$, the above
calculations do not account for the negative integrated weight $W'_{\rm
exp}$ observed around $n \approx 0$ (Fig.\ \ref{f-x}), which has been
attributed in Ref. \cite{Tang} to the onset of the $E_g$ mode in the
presence of a finite potential asymmetry $\Delta$.  These observations
suggest thus that different phonon modes can be optically relevant in
different regions of the phase space.

Our theoretical framework allows us to investigate the possibility of such
phonon switching by taking into account explicitly the role of $\Delta$.
When $\Delta\neq 0$ two effects must be taken into account: ($i$) the
phonon eigenmodes of the systems do not correspond any more to $E_u$ and
$E_g$, even though the new phonon eigenfreqencies $\omega_\pm$ follow
closely the doping dependence of the
$\omega_{\rm A,S}$ of the uncoupled modes \cite{Ando_09,Mauri}. As a
consequence the phonon propagator $D_{\rm AA}$ can develop at large
$\Delta$ a second peak with weaker intensity at approximately the frequency
$\omega_{\rm S}$ of the $E_g$ mode.  Note however that according to
Eq. (\ref{chirice}) the infrared activity of the $D_{\rm AA}$ phonon
propagator is still ruled by the strength $p_{\rm A}$.  Since, as we shall
see below, $p_{\rm A}$ vanishes for $n \rightarrow 0$ also when $\Delta
\neq 0$, the appearance of the IR phonon structures at $n \approx 0$
\cite{noi,Tang} cannot be accounted for by the mixing of the modes; ($ii$)
in addition to the previous effect, the presence of a finite $\Delta$ leads
also to a finite mixed response function $\chi_{j{\rm S}}(\omega)= -\langle
{j} V_{\rm S}\rangle_\omega \neq 0$.  Eq. (\ref{chirice}) must be thus
generalized as:
\begin{eqnarray}
\chi_{jj}^{\rm ep}(\omega)
&=&
\chi_{j{\rm A}}(\omega)D_{\rm AA}(\omega)\chi_{{\rm A}^\dagger j}(\omega)
\nonumber\\
&&
+\chi_{j{\rm S}}(\omega)D_{\rm SS}(\omega)\chi_{{\rm S}^\dagger j}(\omega)
\nonumber\\
&&
+\left[
\chi_{j{\rm A}}(\omega)D_{\rm AS}(\omega)\chi_{{\rm S}^\dagger j}(\omega)
+\mbox{h.c.}\right],
\label{ricegap}
\end{eqnarray}
where $\chi_{{\rm S}^\dagger j}(\omega)=\chi_{j {\rm S}}(\omega)$,
and
$D_{\rm \nu \nu'}(\omega)=
-\langle \phi_{\rm \nu} \phi_{\rm \nu'}\rangle_\omega$ are the phonon
propagators, calculated including the 
hybridized phonon self-energy $\chi_{\rm AS}$ for finite
$\Delta$ \cite{Ando_09,Mauri}.
Eq. (\ref{ricegap}) shows
how, due to the $\chi_{j{\rm S}}$ response triggered by the finite $\Delta$,
a {\em direct} coupling channel to the symmetric $E_g$ mode vibrations
(the $D_{\rm SS}$ phonon propagator) appears in the optical\
spectroscopy.
Similar expressions as Eqs. (\ref{chimix})-(\ref{elements})
can be derived for $\chi_{j{\rm S}}$, where
however the coefficients  $C_{j{\rm S},{\bf k}}^{nm}$,
as well as $C_{j{\rm A},{\bf k}}^{nm}$, have a more complex structure
for $\Delta \neq 0$.

\begin{figure}[t]
\centerline{
\includegraphics[scale=0.5,clip=]{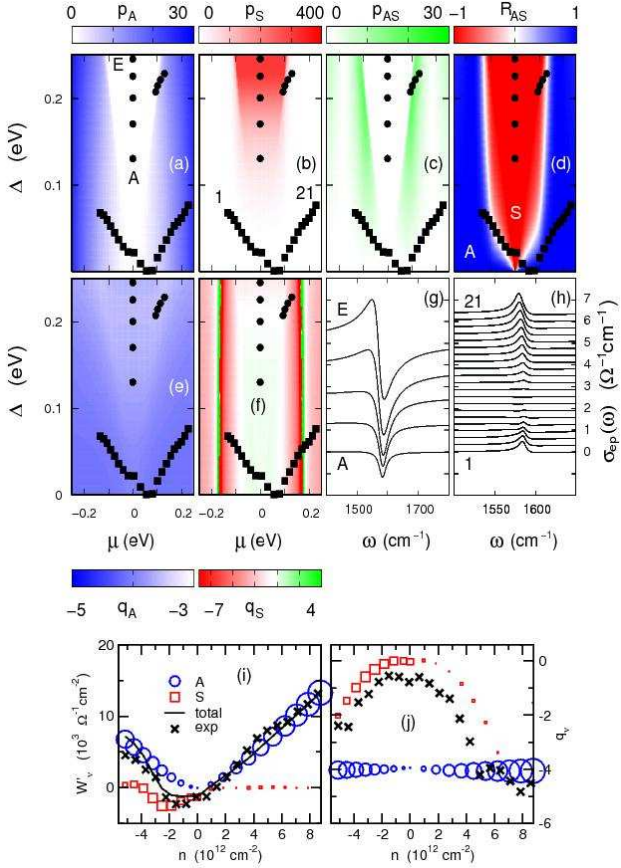}}
\caption{(color online)
(a)-(c) strength $p_\nu$ of the different electron-phonon
contributions to the optical conductivity. The above color scale
is in units of $10^3 \Omega^{-1}$ cm$^{-2}$. (d)
Relative phonon weight $R_{\rm AS}$.
(e)-(f) Fano asymmetry factors $q_{\rm A}$ and $q_{\rm S}$.
Also shown in these plots is the $\Delta$-$\mu$ location
of experimental data: \cite{Tang} ({\Large $\bullet$}), 
\cite{noi} ($\blacksquare$). The corresponding optical spectra are
shown in panels (g) ($\mu=0$ and increasing $\Delta$ from A to E) and (h)
(increasing doping from 1 to 21), 
respectively. Curves are displaced vertically for clarity.  
(i) different electron-phonon contributions $W'_\nu$ (open symbols)
to total $\omega$-integrated spectral weight (solid line) calculated for
$\Delta$-$\mu$ of Ref. \cite{noi}.
(j) corresponding Fano factors $q_\nu$ from Eq. (\ref{fanorice2}).
The size of the open symbols in panels
(i)-(j) is proportional to $p_\nu$.
Crosses are experimental data from Ref. \cite{noi}.}
\label{f-map}
\end{figure}

To elucidate the competition between the different optical channels in
Eq. (\ref{ricegap}) we compute explicitly $\chi_{j{\rm A}}$ and
$\chi_{j{\rm S}}$ for generic $\Delta$ and $\mu$, giving a complete phase
diagram that can be explored in double-gated samples.
To parameterize the
strengths of the relative channels
we use the quantities $p_\nu=\pi|\chi_{j\nu}|^2/\omega_{\rm av}$
for $\nu={\rm A,S}$
and $p_{\rm AS}=\pi|\chi_{j{\rm A}}\chi_{j {\rm S}}| /\omega_{\rm av}$ for the
mixed channel, where $\omega_{\rm av}$ is the average frequency of the two poles
of the phonon propagators \cite{Ando_09,Mauri}.
For a direct comparison with experimental data, we set
$T=10$ K and $\eta=30$ meV, which is halfway between $\eta=18$ meV reported
in Ref. \cite{Kuzmenko2} and $\eta=40$ meV in Ref. \cite{Tang}.
The results for the phonon
strengths and Fano factors are plotted as a color-map in the $\Delta$-$\mu$
space in Fig. \ref{f-map}a-f where also the $\Delta$-$\mu$
location of the available IR data \cite{noi,Tang} are shown.
Fig. \ref{f-map}a,b shows that the main parameter tuning the strength of
the $E_u$ mode is the charge doping whereas the $E_g$ mode is active mainly
around the neutrality point, with a strength that is tuned by the asymmetry
gap $\Delta$.  The strength of the mixed optical structure [third line in
Eq. (\ref{ricegap})], shown in Fig. \ref{f-map}c, satisfies $p_{\rm AS} \ll
p_{\rm A}, p_{\rm S}$ in a large part of the phase diagram so that only the
A and S channels [first two lines in Eq. (\ref{ricegap})] are active.  The
relative intensity $R_{\rm AS}=(p_{\rm A}-p_{\rm S})/(p_{\rm A}+p_{\rm S})$
is summarized in Fig. \ref{f-map}d where $R_{\rm AS}\approx 1$ corresponds
a dominant $E_u$ mode, while $R_{\rm AS}\approx -1$ gives a dominant $E_g$
resonance.  Note that the region $R_{\rm AS} \approx 0$ where the two
strengths of the optical structures have similar magnitude is quite narrow
and difficult to resolve experimentally.  In Fig. \ref{f-map}e,f we plot
also the relative Fano factor $q_\nu$ for the A and S modes. As one can
see, $q_{\rm A}$ is essentially doping independent while $q_{\rm S}$ shows
a sizeable dependence as function of $\mu$. These behaviors can be
understood considering that the structure of Eq. (\ref{chimix}) for
$\chi_{j{\rm A}}$ is still valid for $\Delta \neq 0$, so that the
low-energy $2-3$ interband transitions are missing and the weak Fano
asymmetry of the A mode is only due to a finite broadening due to $\eta$ on
the remaining transitions.  On the contrary {\em all} the interband
transitions contribute to $\chi_{j{\rm S}}$, including in particular the
low-energy $2-3$ interband transitions which overlap with the phonon
frequency (for $2|\mu| \le \omega_{\rm S}$), accounting for the sizable
dependence of $q_{\rm S}$ as function of $\mu$.

The comparison of the present results with the $\Delta$-$\mu$ location of
the experimental available data provides an important route to check our
theoretical predictions. In Fig.\ \ref{f-map}g-h we show the optical
conductivity $\sigma'_{\rm ep}(\omega)$ using Eq.\ (\ref{ricegap}) for the
$\mu,\Delta$ values corresponding to the experimental data of Refs.\
\cite{noi,Tang}. The phonon propagators $D_{\nu \nu'}$ are computed
taking account the self-energy hybridization due to
$\Delta \neq 0$ \cite{Ando_09,Mauri}, and
convoluted with the experimental resolution of 10 cm$^{-1}$ \cite{noi}.  
In Fig.\ \ref{f-map}g we plot the spectra for $\mu=0$, 
showing the evolution of the optical intensity and of the Fano
asymmetry for increasing $\Delta$. These features are related uniquely to the
$E_g$ S mode \cite{Tang}, consistently with panels (a) and (b). On the
contrary the spectra in Fig.\ \ref{f-map}h, evaluated at the doping levels and
electrostatic potentials of Ref. \cite{noi}, are expected to 
show a continuous switching between the 
A and S modes, depending on the values of 
$\mu$ and $\Delta$. To investigate deeper
this issue, we plot in
Fig. \ref{f-map}i the theoretical spectral weights $W'_\nu$ for both the
$E_u$ and $E_g$ modes evaluated on the set of experimental data
$\Delta$-$\mu$ of Refs. \cite{noi}.  The size of the symbols is
proportional to the peak strength $p_\nu$.
The switch between the $E_u$ phonon peak and the $E_u$ one in the regime $n
\in [-1:3] \times 10^{12}$ cm$^{-2}$ is evident and is reflected in a total
spectral weight $W'_{\rm T}=W'_{\rm A}+W'_{\rm S}$ which becomes negative
in such $n$-region, in good agreement with the experimental
$\omega$-integrated weight $W'_{\rm exp}$.  A similar phonon switch is also
evident from the analysis of the Fano asymmetry, reported in
Fig. \ref{f-map}j.  Also here one can distinguish the crossover between a
constant $q$ behavior at large $n$, which we can attribute to the $E_u$
mode, and a drop of $|q|$ at small $n$ in the region where the $E_g$ mode
becomes dominant.  The experimental fit, done with a single mode Fano
formula, presents a similar trend.

We stress that the switching between the $E_u$ and $E_g$ modes in
the optical conductivity discussed here is not related to the possible
appearance at large $\Delta$ of a double-peak structure in each of the
$D_{\rm AA}$ or $D_{\rm SS}$ propagators \cite{Malard,Ando_09,Mauri,Kim}.
Even though the two effects could be present simultaneously at large gap
values, the optical data available so far are outside the region where the
two peaks can be resolved.

In conclusion, in this paper we presented a complete theoretical
description of the phonon resonance in bilayer graphene, that accounts for
both the intensity and the Fano-asymmetry variations as functions of the
density and the gap. We also showed that an optical switching from $E_u$ to
$E_g$ can be induced in a controlled way, providing a full understanding of
the experimental data.

The derived phase diagram for the optical
properties offers a roadmap for the characterization of graphenic systems.
For instance the measurement of the IR intensity of the phonon peak
can provide a useful tool to determine
the doping level in contact-free samples.

We thanks A.V. Balatsky, F. Bernardini, F. Mauri and P. Postorino for
enlightening discussions. This work was supported by the Swiss National
Science Foundation (SNSF) by the grant 200021-120347 and under MaNEP, and
by Italian MIUR project PRIN 2007FW3MJX.

\end{document}